\documentstyle[12pt,twoside,fleqn,qm97]{article}

\newcommand{\AmS}{{\protect\the\textfont2
  A\kern-.1667em\lower.5ex\hbox{M}\kern-.125emS}}

\title{Hadronic Observables: Theoretical Highlights%
       \thanks{Work supported by BMBF, DFG, and GSI.}} 

\author{Ulrich Heinz\address{Inst. f\"ur Theoretische Physik,
                             Universit\"at Regensburg,
                             D-93040 Regensburg, Germany\\}%
       }

\begin{document}
\maketitle

\begin{abstract}
 {\small \noindent
 I present highlights from the parallel sessions on the theory of
 hadronic observables in $e^+e^-$, hadronic and nuclear reactions.
 }
\end{abstract}

\section{OVERVIEW}
\label{sec0}

The parallel talks which I will cover can be assigned to one or
several of the following three subject areas:

\smallskip

\noindent
1. Thermalization, flow, and source sizes;\\
2. Chiral dynamics and disoriented chiral condensates (DCC's);\\
3. New developments for transport models and event generators.

\smallskip 

Rather than commenting on individual talks in any detail, I will use
these headings to classify the highlights presented at this
conference. Before discussing individual results, however, I will
make some general remarks on thermalization, flow and hadronic
freeze-out in high energy reactions which, from the discussions at the
conference inside and outside the lecture halls, I find appropriate
and, hopefully, clarifying. For lack of space I omit references to
talks given at this conference which can be found elsewhere in the
proceedings, mentioning only the {\sf names of the speakers}.

\section{THERMALIZATION, FLOW, AND FREEZE-OUT}
\label{sec1}

``Thermal'' behaviour can arise in many {\em conceptually different}
ways. In each case the ``temperature'' parameter $T$ has a different
meaning. To avoid confusion it is therefore essential to keep the
different concepts of ``thermalization'' separate and to be very
specific about which concept one refers to in a given situation. 

For us the two most important variants of ``thermal'' behaviour are the
following: 

{\bf (1)} The {\em statistical occupation of hadronic phase space} with
minimum information. The ``information'' in this case is provided by
external constraints on the total available energy $E$, baryon number
$B$, strangeness $S$ and, possibly, a constraint $\lambda_s$ on the
fraction of strange hadrons. This leads to ``thermal'' behaviour via
the Maximum Entropy Principle in which the ``temperature'' $T$ and
``fugacities'' $e^{\mu_b/T}$, $e^{\mu_s/T}$ (which in the canonical
approach are replaced by so-called ``chemical factors''
\cite{becattini,BH97}) arise as Lagrange multipliers to implement the
constraints. Examples are nucleon emission from an evaporating
compound nucleus in low-energy nuclear physics and hadronization in
$e^+e^-$, $pp$ and $p\bar p$ collisions (hadron yields
\cite{becattini,BH97} and $m_\perp$-spectra~\cite{sollfrank}). The
number of parameters to fit the data in such a situation is equal 
to the number of ``conserved quantities'' (constraints), and it
reflects directly the information content of the fitted observable(s).
This type of ``thermal'' behaviour requires {\em no} rescattering and
{\em no} interactions among the hadrons, there is no pressure and no
collective flow in the hadronic final state and, in fact, the concept
of {\em local} equilibrium can {\em not} be applied. In other words,
this type of ``thermal'' behaviour is not what we are interested in in
heavy ion collisions, except as a baseline against which to
differentiate interesting phenomena. 

{\bf (2)} Thermalization of a non-equilibrium initial state by {\em
  kinetic equilibration} (rescattering). This does require (strong!)
interactions among the hadrons. Here one must differentiate between
{\em thermal} equilibration (reflected in the shape of the momentum
spectra), which defines the temperature $T$, and {\em chemical}
equilibration (reflected in the particle yields and ratios) which
defines the chemical potentials in a grand canonical description. The
first is driven by the {\em total} hadron-hadron cross section while
the second relies on usually much smaller {\em inelastic} cross
sections and thus happens more slowly. This type of equilibrium is
accompanied by pressure which drives collective flow (radial expansion
into the vacuum as well as directed flow in non-central collisions). 
In heavy ion collisions it is realized {\em at most locally}, in the
form of local thermal and/or chemical equilibrium -- due to the
absence of confining walls there is never global equilibrium. This
type of ``thermal'' behaviour is what we are searching for in heavy
ion collisions.   

I stress that {\em flow} is an unavoidable consequence of this type of
equilibration. Thermal fits without flow to hadron spectra are not
consistent with the kinetic thermalization hypothesis. Flow contains
information; it is described by three additional fit parameters
$\bbox{v}(x)$. This information is related to the pressure history in
the early stages of the collision and thereby (somewhat indirectly) to
the equation of state of the hot matter. 

Most thermal fits work with global parameters $T$ and $\mu$ which, at
first sight, appears inconsistent with what I just said. But here the
role of freeze-out becomes important: freeze-out cuts off the
hydrodynamical evolution of the thermalized region via a kinetic
freeze-out criterium \cite{freeze} which involves the particle
densities, cross sections and expansion rate. In practice freeze-out
may occur at nearly the same temperature everywhere \cite{freeze}. 

Clearly a thermal fit to hadron production data (if it works) is not
the end, but rather the beginning of our understanding. One
must still check the {\em dynamical consistency} of the fit
parameters $T_f$, $\mu_f$, $\bbox{v}_f$: can one find equations of
state and initial conditions which yield such freeze-out parameters?
Which dynamical models can be excluded?

\subsection{Chemical equilibrium analysis of $e^+e^-$, $pp$, and $AA$
  collisions}
\label{sec1a}

In spite of what I said about case {\bf (1)} above, a ``thermal''
analysis of hadron yields in elementary collisions is still
interesting. The interest arises {\em a posteriori} from the observed
universality of the fit parameters, namely a universal
``hadronization'' or ``chemical freeze-out'' temperature $T_{\rm
  chem} = T_{\rm had} \approx 170$ MeV (numerically equal to the old
Hagedorn temperature $T_{\rm H}$ and consistent with the
inverse slope parameter of the $m_T$-spectra in $pp$ collisions
\cite{sollfrank}) and a universal strangeness fraction
$\lambda_s{\approx}0.2{-}0.25$, almost independent of $\sqrt{s}$
\cite{becattini,BH97,BGS97}. This is most easily
understood~\cite{BH97} in terms of a universal critical energy density
$\epsilon_{\rm crit}$ for hadronization which, via the Maximum Entropy
Principle, is parametrized by a universal ``hadronization
temperature'' $T_{\rm had}$ and which, according to Hagedorn, 
characterizes the upper limit of hadronic phase space. Supporting 
evidence comes from the observed increase with $\sqrt{s}$ of the
fitted fireball volume $V_f$ (which accomodates the increasing 
multiplicities and widths of the rapidity distributions). Although
higher collision energies result in larger {\em initial} energy
densities $\epsilon_0$, the collision zone subsequently undergoes more
(mostly longitudinal, not necessarily hydrodynamical) expansion until
$\epsilon_{\rm crit}$ is reached and hadron formation can proceed. The
systematics of the data can only be understood if hadron formation at
$\epsilon{>}\epsilon_{\rm crit}$ (i.e. $T{>}T_{\rm H}$ for the
corresponding Lagrange multipliers) is impossible. With this
interpretation, the chemical analysis of $e^+e^-$, $pp$ and $p\bar p$
collisions does provide one point in the $T$-$\mu_b$ phase diagram (see
Fig.1). -- The only ``childhood memory'' of the collision system is
reflected in the low value of $\lambda_s$, indicating suppressed 
strange quark production (relative to $u$ and $d$ quarks) in the early
pre-hadronic stages of the collision. 

\vspace*{10cm}
\includegraphics{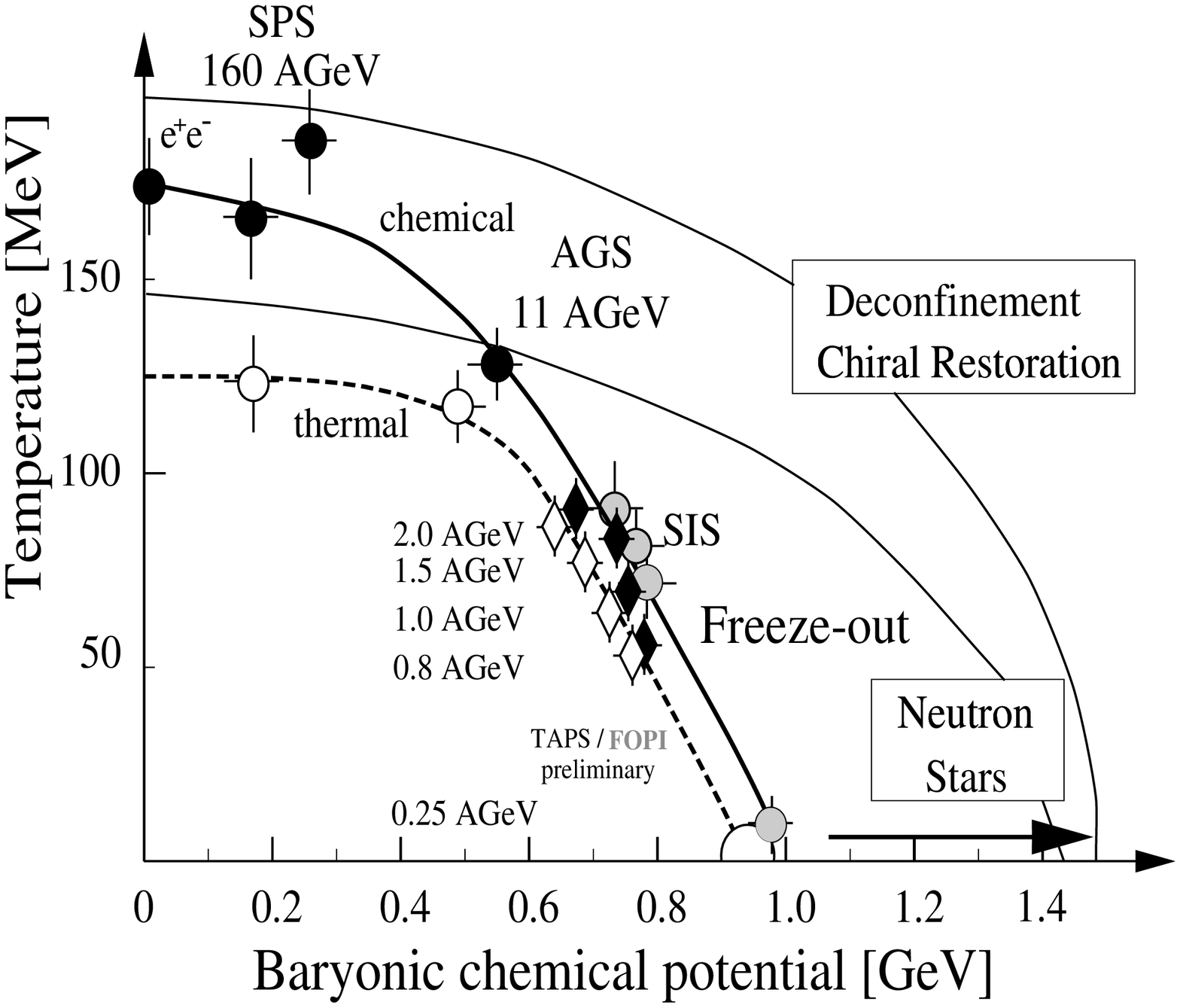}
\begin{center}
\begin{minipage}[t]{13cm}
{\small \noindent {\bf Fig.1.}  
Compilation of freezeout points from SIS to SPS energies. Filled
symbols: chemical freeze-out points from hadron abundances. Open
symbols: thermal freeze-out points from momentum spectra and
two-particle correlations. (For each system, chemical and thermal
freeze-out were assumed to occur at the same value $\mu_B/T$.) The
chemical freeze-out point from $e^+e^-$ collisions
\protect\cite{becattini} has been included while those from $pp$ and
$p\bar p$ collisions \protect\cite{BH97} were omitted for clarity. 
(Generalization of the figure presented by {\sf Braun-Munzinger}
and {\sf Metag} to whom I am grateful for help.)}  
\end{minipage}
\end{center}

In this light the observation \cite{S97} of a chemical freeze-out
temperature $T_{\rm chem} \approx T_{\rm H} \approx 170$ MeV 
in heavy ion collisions at the SPS (Fig.1) is not really
interesting. It suggests not only to the sceptic that in heavy ion
collisions hadronization occurs via the same statistical hadronic
phase space occupation process as in $pp$ collisions. What {\em is}
interesting, however, is the observation ({\sf Becattini})
that the strangeness fraction $\lambda_s{\approx}0.4{-}0.45$ in $AA$
collisions is about a factor 2 larger than in $e^+e^-$ and $pp$
collisions. If $pp$ and $AA$ collisions hadronize via the same
mechanism, and in $AA$ collisions the Maximum Entropy particle yields
fixed at $T_{\rm had}$ are not modified by inelastic hadronic final
state rescattering, this increase in $\lambda_s$ must reflect a {\em
  difference} in the properties of the {\em prehadronic} state! In
nuclear collisions the prehadronic stage allows for more strangeness
production, most likely due to a longer lifetime before hadronization. 

{\sf Sollfrank} showed that $\lambda_s=0.45$ corresponds to a
strangeness saturation coefficient $\gamma_s\approx 0.7$, and 
that the factor 2 rise of $\lambda_s$ in $AA$ collisions cannot be
explained by the removal of canonical constraints on strangeness
production in the small $e^+e^-$ and $pp$ collision volumes. He also
argued that a strangeness saturation of $\gamma_s \approx 0.7$ in the
hadronic final state may be the upper limit reachable in heavy ion
collisions because the corresponding strangeness fraction agrees with
that in a fully equilibrated QGP at $T_{\rm had} \approx 170$ MeV. If
both strangeness and entropy are conserved or increase similarly
during hadronization, $\gamma_s\approx 0.7$ in the Maximum Entropy
particle yield of the final state hadrons would be a universal
consequence of a fully thermally and chemically equilibrated QGP
before hadronization (and the SPS data would be consistent with such a
state)! 

According to Fig.~1 chemical freeze-out at the SPS (and also
at the AGS?) appears to occur right at the critical line, i.e. at
hadronization, whereas the SIS data indicate much lower chemical
freeze-out temperatures ({\sf Metag}). The origin of this is not yet
clear but likely due to longer lifetimes of the reaction zone,
especially at lower beam energies, allowing for chemical equilibration
by inelastic hadronic reactions. 

\subsection{Thermal equilibrium and flow}
\label{sec1b}
 
The other interesting observation in the hadronic sector of nuclear
collisions is that of collective flow (radial expansion flow, directed
and elliptical flow). It is usually extracted from the shape of the
single-particle momentum distributions. Radial flow, for example,
leads to a flattening of the $m_\perp$-spectra. For the analysis one
must distinguish two domains. In the relativistic domain
$p_\perp{\gg}m_0$ the inverse slope $T_{\rm app}$ of all particle
species is the same and given by the blueshift formula \cite{freeze}
$T_{\rm app}{=}T_f \sqrt{(1{+}\langle v_\perp\rangle)/(1{-}\langle
  v_\perp\rangle)}$. This formula does not allow to disentangle the
average radial flow velocity $\langle v_\perp\rangle$ and freeze-out
temperature $T_f$. In the non-relativistic domain $p_\perp{\ll}m_0$
the inverse slope is given approximately by $T_{\rm app}{=}T_f{+}m_0
\langle v_\perp^2 \rangle$, and the rest mass dependence of the
``apparent temperature'' (inverse slope) allows to determine $T_f$ and
$\langle v_\perp^2 \rangle$ separately. (In $pp$ collisions no
$m_0$-dependence of $T_{\rm app}$ is seen \cite{NuXu}.) Plots of
$T_{\rm app}$ against $m_0$ were shown in many talks at this
conference, showing that the data follow very nicely this systematics,
from SIS to SPS energies (open symbols in Fig.~1). Notable (but not
understood) exceptions were the $\Xi$- and $\Omega$-spectra of WA97
({\sf Kr\'alik}) which are steeper than expected from this formula.

\subsection{Rescattering -- yes or no?}
\label{sec1c}

In his overview of the beam energy dependence of flow phenomena {\sf
  Ollitrault} showed that all three types of flow appear in the 
data simultaneously, pointing to rescattering among the secondary
hadrons as a common origin. The difference between the chemical and
thermal freeze-out points in Fig.~1 suggests significant {\em elastic}
rescattering between hadronization and decoupling, causing expansion
and cooling of the momentum distributions. (Elastic collisions
include resonance channels like $\pi{+}N{\to}\Delta{\to}\pi{+}N$ which
do not change particle abundances.) While present SPS data are
consistent with a common chemical freeze-out temperature in small
(S+S) and large (Pb+Pb) collision systems (suggesting that particle
abundances decouple directly after hadronization), thermal freeze-out
seems to happen at lower temperature in Pb+Pb (120-130 MeV) than in
S+S (140-150 MeV). This is consistent with hydrodynamical simulations
({\sf Shuryak}) which show that larger systems live longer, develop
more collectivity and cool down further before breaking apart. Low
thermal freeze-out and large transverse flow in Pb+Pb are confirmed
directly by HBT analyses.  

I find the evidence presented at this meeting for the presence of all
3 types of collective flow at AGS and SPS energies convincing. Still,
it is gratifying that the alternative view, namely that the nuclear
broadening of the $p_\perp$-spectra can be understood without flow
in terms of initial state scattering only \cite{rwm}, can be rejected
by independent methods\footnote{Please note that initial and final
  state scattering effects on the spectra are {\em not additive}; if
  strong enough, both effects together dissolve into collective
  flow. It is therefore not correct to ``subtract'' initial 
  state scattering effects from the slope parameters in order to
  isolate the flow contribution.}. While hadronic single-particle
spectra were shown to have limited discriminating power
\cite{rwmflow}, initial and final state rescattering effects can be
clearly differentiated by studying two-particle HBT correlations and
event-by-event fluctuations. An HBT analysis \cite{THP97} of the
models presented in \cite{rwm} showed that they cannot account for the
observed transverse expansion of the reaction zone, giving an
$R_\perp$ which is factor 2 too small compared to the data, nor for
the radial flow reflected in the observed significant
$M_\perp$-dependence of $R_\perp$, nor the observed growth of
$R_\parallel$ with $A$ which reflects the longer total reaction times
until freeze-out in the larger nuclear collision systems. A study of
event-by-event fluctuations of the average $\langle p_\perp \rangle$
for pions \cite{GLR97} shows that initial state scattering effects
generically increase those fluctuations while the data show a strong
decrease from $pp$ to Pb+Pb, consistent with URQMD rescattering
simulations presented by {\sf Bleicher}. Thus strong (elastic)
rescattering is required by the HBT and fluctuation data. 

\subsection{The power of HBT}
\label{sec1d}

{\sf Pratt, Wiedemann} and {\sf Schmidt-S\o rensen} discussed the
usefulness of two- and three-particle correlations, in conjunction with
single-particle spectra, for a direct reconstruction of the geometry
and dynamical state of the reaction zone at freeze-out. The slopes
of both the single-particle $m_\perp$-spectra and the function
$R_\perp(M_\perp)$ (the transverse HBT radius) are given by
combinations of $T_f$ and $v_f$, the temperature and transverse flow
velocity at freeze-out. But since the two combinations are essentially
orthogonal on each other in the $T_f$-$v_f$ plane, together they allow
to separate thermal from collective motion ({\sf Wiedemann, Roland}). 
For central Pb+Pb collisions at the SPS the average transverse flow
velocity extracted in this way is large, $\langle v_\perp \rangle
\simeq 0.5c$, while the thermal freeze-out temperature is low,
$T_f{\simeq}120{-}130$ MeV. The large transverse flow is accompanied by
a large transverse expansion: the pion source at freeze-out is more
than twice as large as the colliding Pb nuclei. This is quantitatively
confirmed by an analysis of Coulomb effects of the fireball on the
shape of the charged pion spectra at small $p_T$ ({\sf Heiselberg}). 
The source expands rapidly longitudinally, with a nearly
boost-invariant longitudinal velocity profile, for about 9 fm/$c$
before the pions decouple and are emitted over a period of about 2-3
fm/$c$ ({\sf Wiedemann}).  

The pion-emitting source seems to be ``transparent'' rather than
opaque, emitting pions from everywhere, not only from a thin surface
layer. Opacity leads to a smaller outward than sideward HBT radius at
low transverse pair momenta ({\sf Heiselberg}) or a negative
``temporal'' radius parameter in the YKP parametrization
\cite{TH97}. This is excluded by the NA49 data \cite{roland}, while
NA44 data may still allow for $R_{\rm out}^2 < R_{\rm side}^2$ 
at low $K_\perp$ ({\sf Heiselberg}); this requires clarification.   

These studies bring us close to a quantitative characterization
of the final state in heavy ion collisions in {\em phase}-space,
including its geometric and dynamical space-time structure. This
can be used for a strongly constrained {\em extrapolation backward
in time}. In \cite{rio} I presented a semi-quantitative attempt to do so,
going as far back in time as neccessary to shrink the reaction zone 
to its initial size before {\em transverse} expansion (about 1.5 fm/$c$
after impact). Using energy conservation I estimate an initial
energy density of about $\epsilon_0 \approx 2$ GeV/fm$^3 \geq 2
\epsilon_{\rm crit}$. This again points towards a non-hadronic initial
state, with enough local equilibration to drive transverse collective
expansion by thermodynamic pressure\footnote{Whether the observed
  transverse flow is created before or after hadronization cannot be
  decided yet.}.

{\sf Schmidt-S\o rensen} presented the first serious attempt to
extract a true 3-pion correlation signal from heavy-ion collisions. 
For chaotic sources the normalized true 3-pion correlator $r_3$ can be
written as $2\cos\Phi$ where $\Phi$ is the sum of phases of the three
2-body exchange amplitudes and a function of the relative momenta
$q$. $r_3(q{=}0)$ measures the degree of chaoticity of the source, and
its $q$-dependence measures the source asymmetry around its center
\cite{HZ97}. Within the (large) statistical error bars the data of
{\sf Schmidt-S{\o}rensen} can be fit by the functional form
$2\cos\Phi(q)$, consistent with a completely chaotic source.  

\section{CHIRAL DYNAMICS AND DISORIENTED CHIRAL CONDENSATES}
\label{sec2}

The search for DCC's continues to motivate theoretical work to predict
their evolution and experimental signatures. Most existing work
concentrates on the dynamics of the chiral field itself, neglecting
interactions with other types of hadrons in the reaction zone, 
e.g. baryons. One usually tries to solve directly the relativistic
field equations for the chiral field, but such an approach becomes
impractical once interactions with other fields are included. These
are more easily implemented in terms of semiclassical transport models
for test particles or wavepackets. 

{\sf M.~Bleicher} showed a URQMD simulation for Pb+Pb collisions at
the SPS in which a DCC was put in by hand late in the reaction and
seen to be destroyed by subsequent collisions with other hadrons on a
very short time scale of order 1-2 fm/$c$. While this sounds
troublesome (and may be correct) the treatment is not quite
consistent: the DCC itself is not allowed to evolve (and possibly
regenerate) since its dynamics cannot be handled within the existing
transport approach.  

This problem was addressed by {\sf J.~Randrup} who presented an
implementation of the linear $\sigma$ model in form of a
transport code with particles and mean fields. The relativistic field
equations are split into equations for the mean field and the
fluctuations. The latter is Wigner transformed into a transport
equation. Neglecting collisions among the quantum fluctuations, it
takes the form of a generalized Vlasov equation. The numerical
simulation of this set of equations was shown to agree well with a
direct solution of the initial field equations. This new technical
tool for solving the chiral dynamics can now be merged with other
transport codes such that the question raised by {\sf Bleicher} can be
addressed more quantitatively. 

{\sf V.~Koch} argued that the destruction of a DCC by collisions with
thermal pions is not all bad because on the way out the DCC
contributes to the low-mass dilepton spectrum. (This picks out the 
charged pion component of the DCC via the process $\pi^+\pi^- \to
l^+l^-$.) He showed mean-field simulations with up to a factor 100
enhancement in the dielectron spectrum at low $p_T$ but a more
a realistic estimate for the magnitude of the effect would require the
inclusion of collisions.  

{\sf M.~Asakawa} said that he was not worried about the failure by
WA98 and others to see any DCC signals because he believes that
central heavy ion collisions are the wrong place to look for DCC's! 
Not only threaten the large multiplicity densities in central
collisions to destroy the DCC's by collisions, but also the
spontaneous domain growth may happen too slowly. Following a
suggestion by Minakata and M\"uller \cite{MM96}, he proposed instead
collisions at non-zero impact parameter because then the DCC's can be
driven by the magnetic field generated by the charge current of the
two colliding nuclei. The latter couples to the neutral component of
the chiral field via the anomaly $\bbox{E}{\cdot}\bbox{B}$ which
mediates the transition $2\gamma \to \pi^0$ and spawns DCC's by
giving them an initial kick. Quantitative predictions based on this
nice idea are not easy due to uncertainties in the initial conditions,
but his results looked quite promising.  

\section{NEW DEVELOPMENTS FOR TRANSPORT MODELS}
\label{sec3}

A common problem for transport models of the phase-space
evolution in high-energy heavy-ion collisions is to get the correct
nuclear stopping power, i.e. the amount of energy degradation
experienced by projectile baryons when passing through a 
nuclear target. A particular difficulty is the extra rapidity loss
connected with the conversion of a projectile nucleon into a leading
hyperon via associated strangeness production. $pA$ data
indicate stronger stopping for leading $\Lambda$'s than for protons,
causing an extra shift of strangeness production towards target
rapidities which cannot be explained in terms of normal baryon
stopping. The existing versions of HIJING and VENUS do not reproduce
this behaviour~\cite{topor}. The same effect may be responsible for
the more central strangeness production in Pb+Pb than S+S collisions,
in particular for the much more centrally peaked $\Lambda$ rapidity
distribution shown by {\sf G.~Roland} which is presently not understood.

The baryon stopping problem was addressed by {\sf K.~Geiger}, {\sf
  Y.~Nara}, and {\sf S.E.~Vance}. {\sf Geiger} presented results from
his code VNI which consists of a perturbative partonic cascade
followed by cluster hadronization. {\sf Nara} implemented rescattering
effects among produced hadrons into HIJING. {\sf Vance} modified
HIJING by including, with 25\% relative probability, a new process
which breaks up the leading diquarks from a nucleon-nucleon collision,
creating a ``baryon junction''. After string breaking these junctions
create baryons near midrapidity. All three suggestions lead 
to enhanced baryon stopping and higher net baryon density near
midrapidity. There are differences in detail which need to be sorted
out. No systematic investigation of the different shapes of proton and
$\Lambda$ rapidity distributions was presented yet which thus remains
an open question.

\section{CONCLUDING REMARKS}
\label{sec4}

The present generation of experiments has come a long way towards the 
goal of measuring as many different experimental observables
simultaneously as possible and correlating them with each other. 
Hence the times are over when theorists could get away with trying to
explain single pieces of data. We must now begin to adopt a global
view of the reaction. The present data allow to combine a variety of
different signatures in a controlled way and thereby test theoretical
models in many corners simultaneously. In this way we can begin
to {\em eliminate} models. Natural selection must be permitted to
work, leaving only the most successful theories in the
competition. How else are we going to find out the truth in this
complex field?   

In the hadronic domain the HBT analysis of two- and three-particle
correlations has been established at this meeting as a powerful and
quite practical new tool for our understanding of heavy-ion
dynamics. I am sure that it will now rapidly show its full potential,
including the interesting generalization of the method to unlike
particle correlations~\cite{misk}.  

A new and quite promising field where clearly much more theoretical
guidance is required is event-by-event physics. The NA49 data
\cite{roland} show that, on the $10^{-3}$ level, all Pb+Pb collisions
are alike. The obvious conclusion is that if QGP is made in Pb+Pb
collisions at the SPS, {\em it is made in every collision}! On the
other hand, not seeing any qualitative structures in the fluctuation
spectra makes their analysis more difficult than originally thought;
a {\em quantitative} understanding of their widths is required.
Theorists are just beginning to take up that challenge, and no final
results were reported yet at this meeting.  
 


\end{document}